\documentclass[
    ,final    ]
  {aipproc}

\layoutstyle{6x9}

\usepackage{bm}

\begin{document}

\newcommand{\be}{\begin{equation}}
\newcommand{\ee}{\end{equation}}
\newcommand{\bea}{\begin{eqnarray}}
\newcommand{\eea}{\end{eqnarray}}
\newcommand{\bfk}{\bm{k}}
\newcommand{\pup}{p^\uparrow}
\newcommand{\qup}{q^\uparrow}
\newcommand{\bfp}{\bm{p}}
\newcommand{\nd}{\noindent}
\newcommand{\la}{\lambda}

\vspace*{-1.cm}

\title{Transverse Momentum Dependent Distributions in Hadronic
  Collisions: \\
$p^\uparrow p\to D+X$ and $p^\uparrow p\to \gamma +X$\footnote{Talk
  delivered by U.~D'Alesio at the ``17th International Spin Physics
  Symposium'', SPIN2006, October 2-7, 2006, Kyoto, Japan.}}



\author{U. D'Alesio}{
  address={Dipartimento di Fisica, Universit\`a di Cagliari and \\
INFN, Sezione di Cagliari, C.P. 170, I-09042 Monserrato (CA),
Italy}
}

\author{F. Murgia}{
  address={Dipartimento di Fisica, Universit\`a di Cagliari and \\
INFN, Sezione di Cagliari, C.P. 170, I-09042 Monserrato (CA),
Italy}
}

\begin{abstract}
Our understanding of the transverse spin structure of hadrons might definitely get improved
by the information we gather on transverse momentum dependent (TMD) distributions.
These new functions could also be crucial for a description of the observed transverse single
spin asymmetries (SSA). In a hard scattering model for inclusive hadronic reactions, based on
a generalized QCD factorization scheme, many mechanisms - namely the Sivers \cite{siv},
Collins \cite{col}, Boer-Mulders \cite{dan} effects - might contribute to a SSA. We show how
the $\bfk_\perp$ dependent phases arising from the partonic kinematics together with a
suitable choice of experimental configurations could help in disentangling the above
mentioned effects. We discuss their potential role in two inclusive hadronic processes: heavy
meson and photon production in $pp$ and $p\bar{p}$ collisions.
\end{abstract}

\maketitle


\section{Introduction and Formalism}
\label{intr}

The study of transverse single spin asymmetries and their interpretation in terms of parton
degrees of freedom could open a new window in our understanding of the internal spin
structure of hadrons. In collinear pQCD at leading twist SSA are almost vanishing, at
variance with the large values observed for instance in $p^\uparrow p\to\pi X$ processes.
Among the different approaches proposed in the literature, we consider here a pQCD
generalized factorization scheme with inclusion of intrinsic transverse momentum effects. At
this stage this factorized description of $A \, B \to C \, X$ processes has to be regarded as
a phenomenological model based on a natural extension of the usual collinear approach for the
same process. The factorized scheme with unintegrated partonic distributions has been
recently proven for SIDIS and Drell-Yan processes \cite{jmy}. In a series of
papers~\cite{fu,noi-1,noi-2} we have shown how, within the helicity formalism, a careful
treatment of the noncollinear partonic configurations lead to the appearance of several spin
and TMD parton distribution (pdf) and fragmentation functions (ff) together with a complex
structure in terms of $\bfk_\perp$ dependent phases. Schematically, for the (single)
polarized cross section we have (see Ref.~\cite{noi-2} for details) \bea d\sigma^{A,S_A + B
\to C + X} = && \hspace*{-1.3cm}\sum_{a,b,c,d, \{\la\}} \!\!\!\!
\rho_{\la^{\,}_a,\la^{\prime}_a}^{a/A,S_A} \, \hat f_{a/A,S_A}(x_a,\bfk_{\perp a})\!
\otimes\! \rho_{\la^{\,}_b, \la^{\prime}_b}^{b/B} \,
\hat f_{b/B}(x_b,\bfk_{\perp b})  \nonumber \\
&\otimes & \!\!
\hat M_{\la^{\,}_c, \la^{\,}_d; \la^{\,}_a,\la^{\,}_b}
\,\hat M^*_{\la^{\prime}_c, \la^{\,}_d;
  \la^{\prime}_a,\la^{\prime}_b} \otimes
\hat{D}^{\la^{\,}_C,\la^{\,}_C}_{\la^{\,}_c,\la^{\prime}_c}
(z,\bfk_{\perp C}) \,.
\eea
One of the difficulties in gathering experimental information on these new spin and
$\bfk_\perp$ dependent pdf's and ff's is that most often two or more of them  contribute to
the same physical process, making it very difficult to estimate each single one
separately. In Refs.~\cite{noi-1,noi-2} it was explicitly shown that for the transverse SSA,
$A_N$, for inclusive pion production, in the kinematical region of large positive $x_F$, the
only sizeable contributions come from the Sivers and, less importantly, from the Collins
mechanisms.

It is then worth to consider other inclusive processes in various kinematical configurations,
in order to be sensitive more directly to one particular mechanism. To this aim a careful
choice of the final state could already simplify the task by reducing the partonic
subprocesses and therefore the possible mixing up of different effects. We will then discuss
here SSA for the inclusive production of $D$ mesons and photons in $pp$ (and $p\bar p$)
collisions and show how such a strategy could be carried out. To this end we will employ
maximized TMD distribution and fragmentation functions, keeping however their proper
azimuthal phases. Namely, we adopt for each spin and TMD distribution its trivial positivity
bound.

\vspace*{-.5cm}

\section{SSA in $pp\to D + X$}
$D$ mesons originate predominantly from $c$ or $\bar c$ quarks, which at LO can be created
either via $q \bar q$ annihilation, $q \bar q \to c \bar c$, or via a gluon fusion process,
$gg \to c \bar c$.

As the gluons cannot carry any transverse spin the elementary process $gg \to c \bar c$
results in unpolarized final quarks. In the $q \bar q \to c \bar c$  process one of the
initial (massless) partons, that inside the transversely polarized proton, can be polarized;
however, there is no single spin transfer in this $s$-channel interaction so that again the
final $c$ and $\bar c$ cannot be polarized (no Collins effect). Analogously, for an
$s$-channel process no Boer-Mulders effect can be active. We have explicitly verified that
all contributions to $A_N(\pup p \to D\;X)$ from $\bfk_\perp$ dependent pdf's and ff's, aside
from those involving the Sivers functions, $\Delta^N f_{a/p^\uparrow}$ or $f_{1T}^{\perp}$
(see Ref.~\cite{trento} for notation), enter with phase factors which make the integrals over
the transverse momenta negligibly small.

In a former paper \cite{noi-3} it was shown how at RHIC energies, $\sqrt{s} = 200$ GeV, gluon
fusion dominates the whole $\pup p \to D\,X$ process, up to $x_F \simeq 0.6$, allowing a direct
access to the gluon Sivers function (GSF). Here we extend that analysis at lower energies,
like those reachable at J-PARC, $\sqrt s =$ 10 GeV, (left panel in Fig.~1), and at the
proposed PAX experiment at GSI, $\sqrt s=$ 14 GeV, (Fig.~1, right panel). Clearly the $q\bar
q$ annihilation process becomes dominant now, giving the opportunity to  access directly the
quark Sivers function (QSF). Notice how with $p\bar p$ collisions at PAX the potential QSF
dominance is even more dramatic. A more detailed study will be presented
elsewhere~\cite{noi-D2}.
\begin{center}
\begin{figure}[t]
\includegraphics[height=.3\textheight,angle=-90]{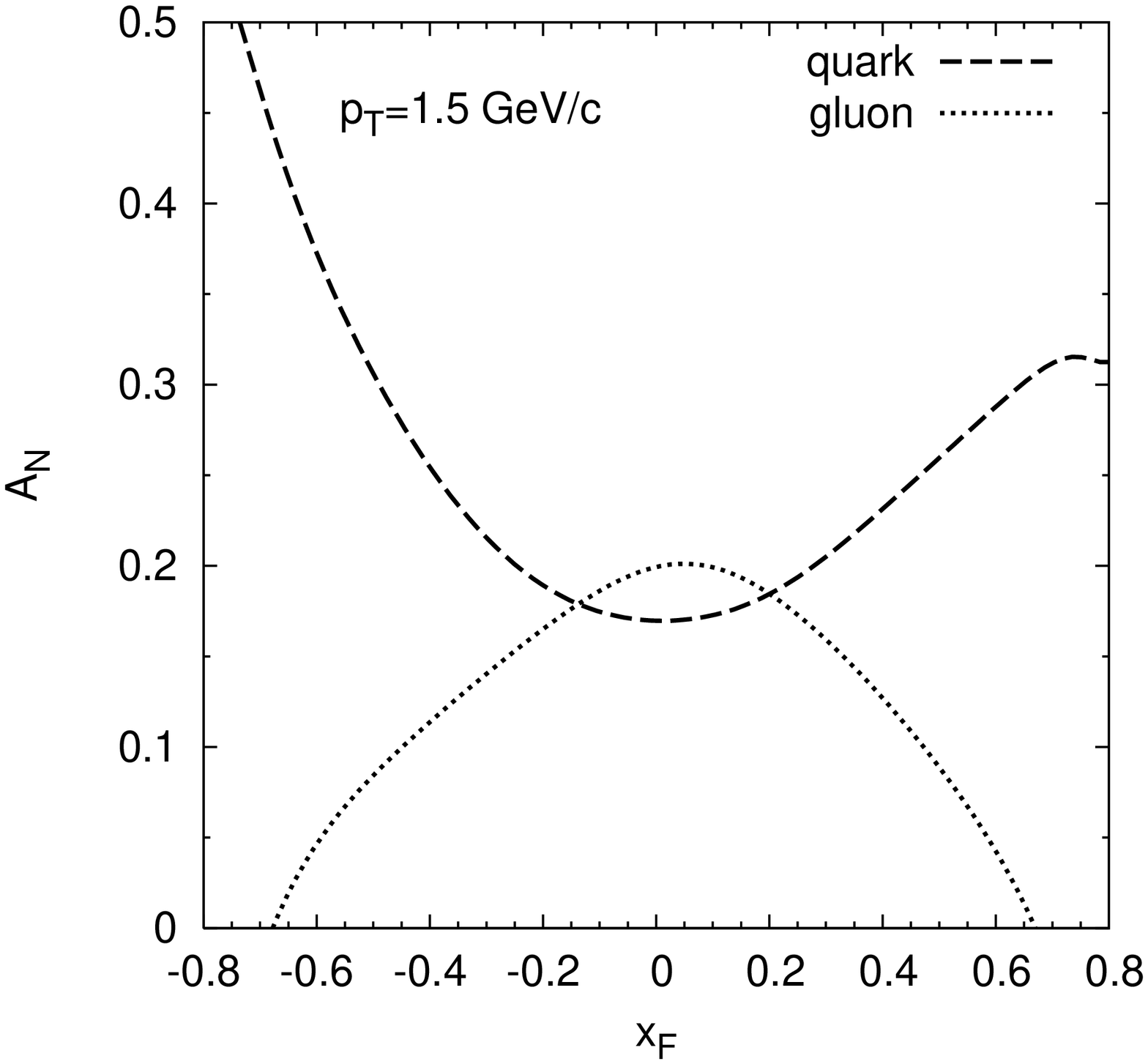}
\includegraphics[height=.3\textheight,angle=-90]{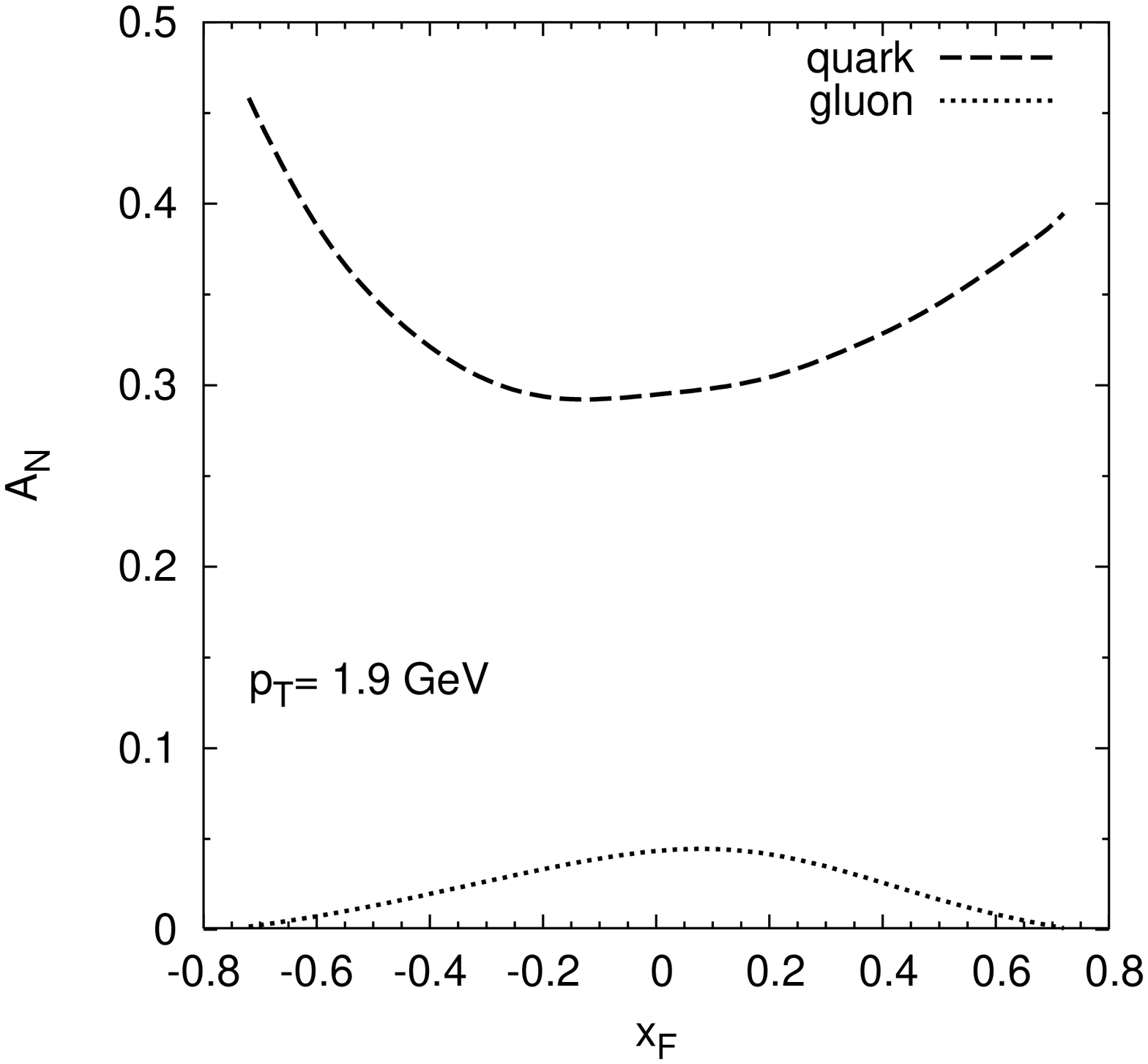}
\caption{Preliminary results for the maximized Sivers contributions (quark, gluon) to $A_N$,
for $pp\to D\,X$ at J-PARC, $\sqrt s= 10$ GeV, (left) and
  for $p\bar p \to D\, X$ at PAX, $\sqrt s= 14$ GeV (right).
}
\end{figure}
\end{center}

\section{SSA in $pp\to \gamma + X$}
The inclusive production of direct photons in $pp$ collisions is certainly a useful tool to
access TMD pdf's, due to the absence of any fragmentation process. Here again we have to
consider only two partonic subprocesses: $q(\bar{q})g\to\gamma\; q(\bar{q})$ and $q\bar q\to
\gamma\;g$. In principle there are three mechanisms that could contribute to the SSA: the
quark Sivers effect, entering both subprocesses, the gluon Sivers effect via the Compton-like
subprocess, and the Boer-Mulders effect (coupled with the transversity pdf, $h_1$) via $q\bar
q$ annihilation. Notice that the electromagnetic coupling will enhance the $u$-flavor
contributions. For $x_F>0$, at fixed $p_T$, the dominant contribution to $A_N$ comes from the
quark Sivers effect. Adopting the parameterizations of the QSF extracted from the fit to
$A_N(p^{\uparrow}p\to\pi\;X)$, one gets a positive SSA, rising with $x_F$~\cite{fu}. Notice
that the collinear twist-3 approach developed by Qiu and Sterman~\cite{qs} and reassessed
recently~\cite{tw3-new} leads to a similar description of $A_N$ for pion production, while
$A_N$ for photon production comes out with an opposite sign.

Concerning the backward rapidity region, it was qualitatively argued~\cite{sof} that at RHIC
energies and at large $p_T$ (around 20 GeV) $A_N$ would be sensitive directly to the GSF. In
our more quantitative approach, with proper treatment of the noncollinear kinematics and the
relative phases, we find that the best region to hunt for the GSF at RHIC energies is, at
$x_F<0$, at $p_T$ values around 5-8 GeV. In this region $A_N$ from the GSF could be as large
as 10\% whereas the other mechanisms would give at most a 1\% contribution.

An even more interesting case is the SSA at lower energies, like for instance at the J-PARC
and PAX experiments. In the first case the QSF gives the main contribution in the forward as
well as in the central rapidity region, whereas again the negative $x_F$ region is dominated
by the GSF (dotted line in Fig.~2, left). At PAX, by colliding polarized protons against
unpolarized antiprotons another effect becomes accessible: namely the Boer-Mulders function,
$\Delta^{N}f_{q^{\uparrow}/p}$ or $h_{1}^{\perp}$, coupled to the transversity distribution.
In this configuration the Compton-like subprocess is suppressed because: $i)$ the minimum $x$
value reached is quite large, being of the order of $2p_T/\sqrt s$; $ii)$ the $\bar q$ in
$\bar p$ has a valence component. We have then a clear dominance of the $q\bar q$ subprocess.
Moreover, the integration over the $\bfk_\perp$-dependent phases does not wash out the
partonic double spin asymmetry. The maximized contribution to $A_N$ from the quark Sivers
effect (dashed line in Fig.~2, right) is still dominating but the one coming from the
Boer-Mulders effect might give $A_N$ values of the order of 20-30\% (dot-dashed line, Fig.~2,
right). By using the parameterizations so far extracted for the QSF one gets $A_N$ of the
order 5-10\% for $p_T$ around 2-4 GeV (solid line, Fig.~2, right). This means that a larger
SSA observed in this region should be a clear signal of the Boer-Mulders effect and then
could be another way to access the transversity distribution. A detailed study is in
progress~\cite{noi-gamma}.

In conclusion, the combined analysis of several inclusive processes, in different kinematical
situations, may provide a strategy for a better determination of the polarized TMD pdf's and
ff's that could be responsible for several large observed azimuthal asymmetries. We have
reported encouraging preliminary results for the inclusive $D$ meson and $\gamma$ production
cases. These results suggest that useful information on the GSF and the Boer-Mulders pdf can
be obtained. A more detailed study is in progress and will be presented
elsewhere~\cite{noi-D2,noi-gamma}.
\begin{center}
\begin{figure}[t]
\includegraphics[height=.3\textheight,angle=-90]{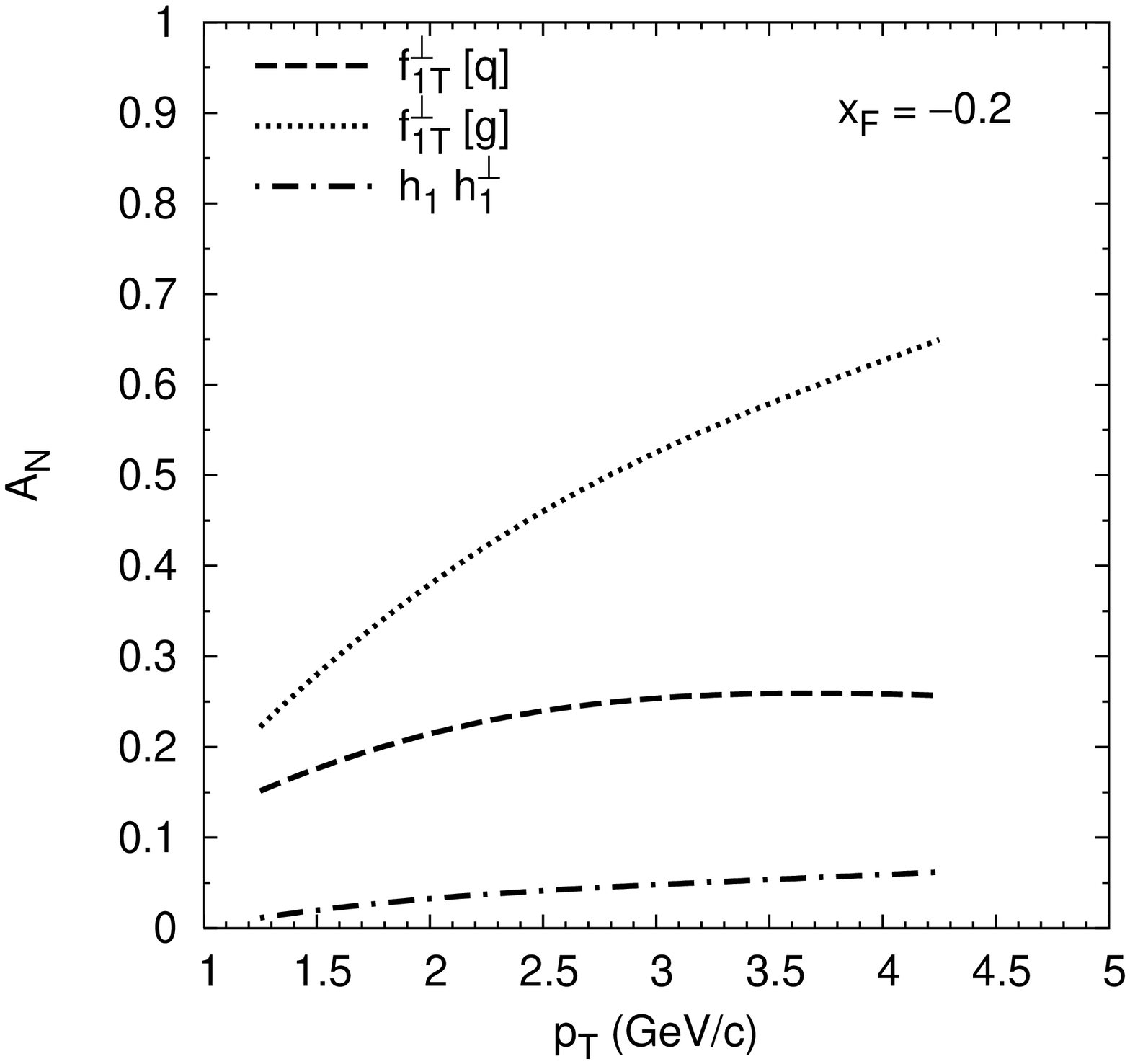}
\includegraphics[height=.3\textheight,angle=-90]{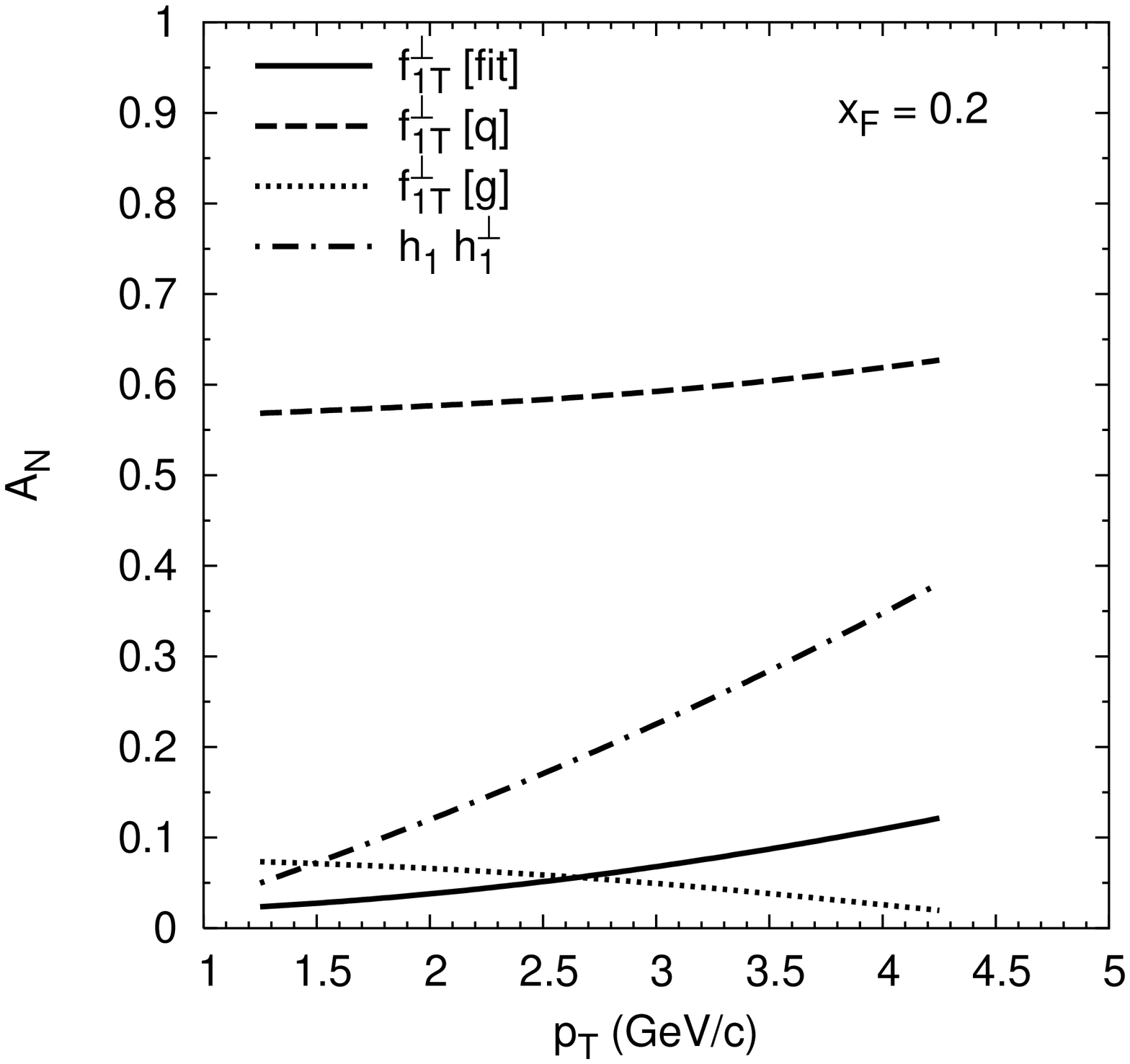}
\caption{Preliminary results for the maximized contributions to $A_N$,
  for $pp\to \gamma\, X$ at J-PARC, $\sqrt s= 10$ GeV and   $x_F=-0.2$
  (left), and for $p\bar p \to \gamma\, X$ at PAX, $\sqrt s= 14$ GeV and
  $x_F=0.2$ (right). 
}
\end{figure}
\end{center}

\bibliographystyle{aipproc}   

\vspace*{-1cm}

\begin{thebibliography}{99}

\bibitem{siv}
  D.W.~Sivers,
  \emph{Phys. Rev.} {\bf D41}, 83 (1990);
  {\bf D43}, 261 (1991).
\bibitem{col}
  J.C.~Collins,
  \emph{Nucl. Phys.} {\bf B396}, 61 (1993).
\bibitem{dan}
  D.~Boer, \emph{Phys. Rev.}  {\bf D60}, 014012 (1999).
 \bibitem{jmy}
   X.D.~Ji, J.P.~Ma, and F.~Yuan, \emph{Phys. Rev.} {\bf D71},
   034005 (2005); \emph{Phys. Lett.} {\bf B597}, 299 (2004).
\bibitem{fu}
  U.~D'Alesio and F.~Murgia,
  \emph{Phys. Rev.} {\bf D70}, 074009 (2004).
\bibitem{noi-1}
  M.~Anselmino, M.~Boglione, U.~D'Alesio, E.~Leader, and F.~Murgia,
  \emph{Phys. Rev.} {\bf D71}, 014002 (2005).
\bibitem{noi-2}
  M.~Anselmino, M.~Boglione, U.~D'Alesio, E.~Leader, S.~Melis, and F.~Murgia,
  \emph{Phys. Rev.} {\bf D73}, 014020 (2006).
\bibitem{trento}
  A.~Bacchetta, U.~D'Alesio, M.~Diehl, and C.~Andy Miller,
  \emph{Phys. Rev.} {\bf D70}, 117504 (2004).
\bibitem{noi-3}
  M.~Anselmino, M.~Boglione, U.~D'Alesio, E.~Leader, and F.~Murgia,
  \emph{Phys. Rev.} {\bf D70}, 074025 (2004).
\bibitem{noi-D2}
  M.~Boglione, U.~D'Alesio, and F.~Murgia, in progress.
\bibitem{qs}
 J.W.~Qiu and G.~Sterman, \emph{Phys. Rev. Lett.} {\bf 67}, 2264
 (1991); \emph{Phys. Rev.} {\bf D59}, 014004 (1999).
\bibitem{tw3-new}
 C.~Kouvaris, J.W.~Qiu, W.~Vogelsang, and F.~Yuan, \texttt{hep-ph/0609238}.
\bibitem{sof}
  I.~Schmidt, J.~Soffer, and J.J.~Yang,
  \emph{Phys. Lett.} {\bf B612}, 258 (2005).
\bibitem{noi-gamma}
  U.~D'Alesio, S. Melis, and F.~Murgia, in progress.

\end{thebibliography}

\end{document}